\documentclass[usenatbib]{mn2e}
\usepackage{graphicx}

\title[Distance estimates of the dark cloud  MBM 12 and the translucent cloud MBM 16]{MBM 12 and MBM 16 distances}
\author[J. Knude and H. E. P. Lindstr{\o}m]{J. Knude $^{1}$\thanks{E-mail:
indus@nbi.ku.dk (JK); heplindstroem@yahoo.dk (HL)} and H. E. P. Lindstr{\o}m$^{1,2}$\footnotemark[1]\\
$^{1}$Niels Bohr Institutet, Copenhagen University, Juliane Maries Vej 30, DK-2100 K{\o}benhavn {\O}, Denmark\\
$^{2}$CSC Danmark A/S, Retortvej 8, DK-2500 Valby, Denmark}
\begin{document}

\date{Accepted 2012 Month 99. Received 2012 Month 99; in original form 2012 Month 99}

\pagerange{\pageref{firstpage}--\pageref{lastpage}} \pubyear{2012}

\maketitle

\label{firstpage}

\begin{abstract}

Among the multitude of intrinsic SDSS index vs. index diagrams the $(g-r) \ vs. \ (r-i)$ diagram is
characterized by showing only minor $(g-r)$ variation for the M dwarfs. The $(g-r) \ vs. \ (r-i)$
reddening vector has a slope almost identical to the slope of the main sequence earlier than $\approx$M2,
meaning that dwarfs later than $\sim$M2 are not contaminated by reddened dwarfs of earlier type.
Chemical composition, stellar activity and evolution have only minor effects on the location of the 
M2$-$M7 dwarfs in the $(g-r) \ vs. \ (r-i)$ diagram implying that reddening may be isolated in a rather 
unique way. From $r$, $M_{r,(r-i)_0}$ and $E_{g-r}$ we may construct distance vs. $A_r$ diagrams.
This purely photometric method is applied on SDSS DR8 data in the MBM 12 region. 
We derive individual stellar distances with a precision $\approx$20$-$26$\%$. For extinctions in the 
$r-band$ the estimate is better than 0.2 mag for $\approx 67 \%$ and between 0.3 and 0.4 for the
remaining $\approx 33 \%$. The extinction discontinuities noticed in the distance vs. $A_r$ diagrams
suggest that MBM 12 is at $\approx$160 pc and MBM 16 at a somewhat smaller distance $\approx$100 pc.
The distance for which $\Delta (A_r)$/$\sigma (\Delta(A_r))$ = 3, where $\Delta (A_r)$ refers to 
$\overline{A_{r, on}}$-$\overline{A_{r, off}}$, may possibly be used as an indicator for the cloud distance.
For MBM 12 and 16 these distance estimates equal 160 and 100 pc, respectively 
\end{abstract}

\begin{keywords}
molecular clouds -- interstellar extinction -- distances : M dwarf stars.
\end{keywords}



\section{Introduction}

MBM 12 is a high latitude molecular cloud and since it has $A_V$ exceeding 5 mag, \citet{jk08}, it is classified as a
dark cloud. And it contains more than a dozen PMS stars. An active high latitude cloud requires possibly
other mechanisms for star formation than dark clouds close to the galactic plane. Estimating parameters 
essential for initiating star formation, such as the cloud mass and density, depends on the cloud distance.
We include MBM 16, a high latitude neighboring tranclucent cloud, only ten degrees removed from MBM 12, and showing
no star formation. For a recent review of high latitude molecular clouds \citet{b_hli_08} may be consulted.

The MBM 12 distance has a long long history and a corresponding, large variation. Shortly after its inclusion
in the MBM catalog \citet{jk02} suggested $\sim$65 pc as derived from NaI spectroscopy and spectroscopic 
distances of a small sample of stars. \citet{jk03} suggest the same distance for MBM 16 about ten degrees
away from MBM 12. The estimate for MBM 12 was altered by \citet{jk08}, after
Hipparcos parallaxes became available, to the range from 58 to 90 pc. Again lower and upper estimates
were based on the absence/presence of NaI absorbtion. \citet{b_hli_05} and \citet{b_hli_06} have used
more indirect methods and prefer a cloud distance of 275 pc and 360$\pm$30 pc, respectively. Both of these
papers do, however, detect dust, thought not to be associated to MBM 12, at 65, 140 pc and at $\sim$80 pc 
respectively. An intermediary distance, 325 pc, is proposed by \citet{b_hli_07} from Vilnius
photometry of dwarfs and giants brighter than V$\approx$12 mag. That work also indicates the possible
presence of a small hump, $A_V \leq$ 0.4 mag, of extinction at 140$-$160 pc.
    
Accurate and homogenous SDSS $griz$ photometry has proven useful for deriving distances and
extinction estimates for stars in the M dwarf range within $\approx$1 kpc \citep[e.g.][]{b_hli_04}. Our
own interest in using $gri$ photometry for 3-D mapping of the ISM was inspired by the stellar models by
\citet{jk04}. These models convinced us that the position of the $(g-r)_0 \ - \ (r-i)_0$ locus
was influenced very little by variation in [M/H], and by ages varying from a few
million years to the age of the Milky Way. The latter invariance was perhaps to be expected. Furthermore,
and most important, the reddening ratio $E_{r-i}/E_{g-r}$ has a value, \citet{jk04}, that
approximates the slope of the main sequence, in the $(g-r) \ - \ (r-i)$ diagram, for dwarfs earlier than
$\sim$M1. A consequence is, that reddening seems to be the only parameter that shifts observed $(g-r)$,
$(r-i)$ pairs away from the intrinsic locus. Another fact that makes the $gri$ photometry so useful for
M dwarfs is the good relation, $\sigma_{M_r} \approx$0.4 mag, between $(r-i)_0$ and the 
absolute magnitude $M_r$: a range of
$\approx$ 2 mag in $(r-i)_0$ corresponds to a range of $\approx$9 mag in $M_r$. This is taken from Table 2
of \citet{jk05}  together with the equation for $M_{r,(r-i)_0}$, Table 4 of \citet{jk09}
valid for the 0.62 $-$ 2.82 range of $(r-i)_0$. The absolute magnitude calibration have required 
precise distances for a represensative sample, Table 4.1, \citet{jk10}.  

We apply the photometric parallax and investigate the variation of extinction with distance in a
region with a radius 15$^{\circ}$ centered on the high latitude, dark cloud MBM 12. This region 
is known to contain several clouds with molecular gas, e.g. the extended translucent cloud MBM 16.
The locations of these clouds are interesting because of their various state of activity and their position
relative to the confinement of the local cavity.       



\section{M Dwarf Sample Selection from SDSS DR8}

\begin{figure} 
\includegraphics[width=88mm]{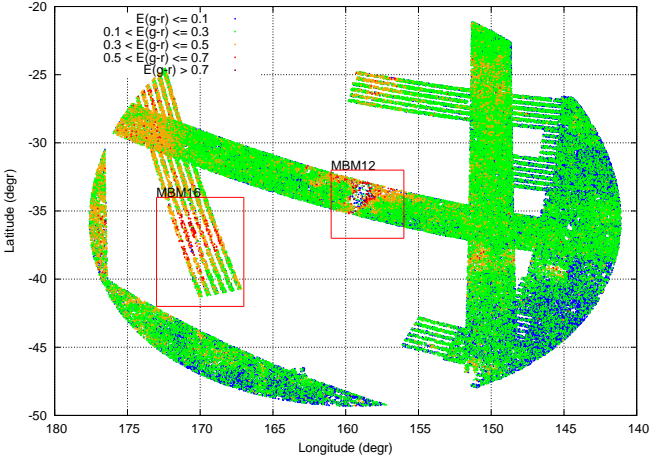}
 \caption{MBM 12 $\underline{region}$. SDSS DR8 M2 - M7 dwarfs with $\sigma_{gri} <$0.060 mag and less than 15$^{\circ}$ 
from the nominal center of MBM 12. The MBM 12 and MBM 16 $\underline{areas}$ area
outlined. The resulting color excesses are color coded} \label{f1}
\end{figure}

\begin{figure} 
\includegraphics[width=80mm]{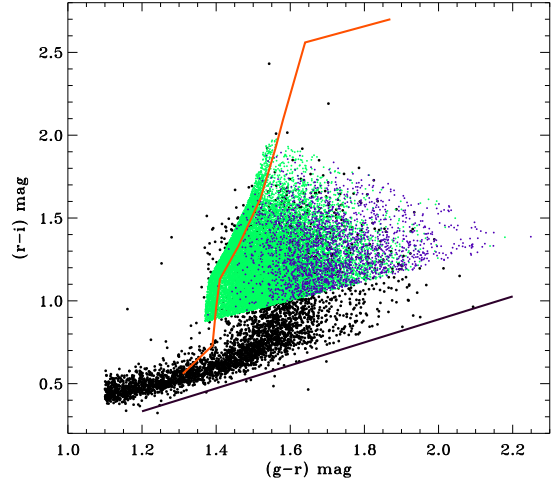}
 \caption{$(r-i) \ vs. \ (g-r)$ diagram. Black points are a selection of stars from
the region in Fig.~\ref{f1}. The green symbols are M dwarfs selected in the MBM 12 area.
Blue symbols are M dwarfs selected in the MBM 16 area. Notice that the earliest M dwarfs
among the MBM 16 stars are shifted to the red of the standard locus which may indicate the presence of
some very local dust.
The solid jagged line is the standard $(g-r)_0 $-$ (r-i)_0$ locus from \citet{jk05} Table 2 given as the
median color of each spectral type $M0,...,M9$. The slope of the cut off equals the 
reddening ratio $E_{r-i}/E_{g-r}$=0.694, Girardi et al. \citet{jk04}, indicated by the straight line} \label{f2}
\end{figure}

To establish a sample of M dwarfs we follow a selection procedure in line
with \citet{b_hli_04}. The candidate sample is drawn from the most recent SDSS data 
release, DR8, applied for a region, centered on the MBM 12 position
$(l,b) \sim (159.^o4, -34.^o3)$ with a 15$^{\circ}$ radius, as shown in Fig.~\ref{f1}. 
The coverage is incomplete but several MBMs are partially scanned.
Using the DR8 CasJobs tool we queried for STAR objects with CLEAN photometry for
$g$, $r$, $i$ and $z$, respectively, and fulfilling the selection criteria:
$\sigma_{g,r,i} <$ 0.06 mag, $(r-i) >$ 0.53 mag and $(i-z) >$ 0.3 mag. The cuts
in $(r-i)$ and $(i-z)$ contribute to minimize the contamination by (partially) eliminating quasars, giants and M flare stars.
Aspects of the sample are shown in Fig.~\ref{f2}. Notice how the reddening 
of the main sequence stars earlier than $\sim$M1 does not contaminate the location of reddened
M dwarfs.

\begin{figure} 
\includegraphics[width=80mm]{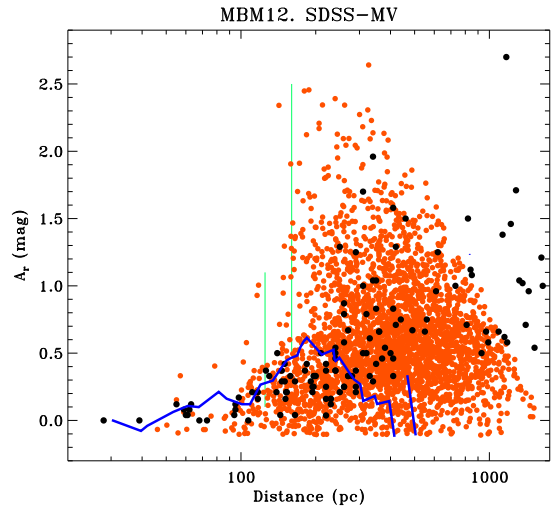}
 \caption{Distance $vs.$ extinction diagram for the MBM 12 area. Red symbols are for the MBM 12 area 
exclusively. The vertical line at 160 pc indicates our suggested MBM 12 location. For comparison the 
black points are the data, with $A_V$ converted to $A_r$, on which a MBM 12 distance of 325 pc was 
based, \citet{b_hli_07}. The blue curve is $\overline{A_r}$ from Hipparcos data} \label{f3}
\end{figure}

\section{Extinction and Distance Estimation. Uncertainty}

The linchpin for our work is the $(r-i) \ vs. \ (g-r)$ diagram, which is
characterized by showing only minor $(g-r)_0$ variation for M dwarfs. The $(r-i) \ vs. \ (g-r)$
reddening vector has a slope almost identical to the slope of the main sequence earlier than $\sim$M1-M2
meaning that reddened dwarfs earlier than this limit do not contaminate the location of reddened dwarfs later
than $\sim$M2. See Fig.~\ref{f2} where the M dwarf standard locus starts at M0.
As judged from the evolutionary models by \citet{jk04} the chemical composition and evolution
have minor effects on the location of the M dwarfs in the $(r-i)_0 \ vs. \ (g-r)_0$ diagram. According
to \citet{jk06} $(g-r)$ colors are likely to depend on metallicity but no clear trends are apparent
for $(r-i)$ and $(g-r)$. M dwarf
activity in the form of flares mostly effects the blue part of the spectrum, i.e. the $u$ and the
$g$ bands implying that $g-r$ is affected but not $r$. According to \citet{jk11} a flare will
typically change $g-r$ to values below 0.05 mag. As Fig.~\ref{f2} shows the $g-r$ range we consider is far
redder than this and $E_{g-r}>$1.4 or $A_r >$4 is required to shift a flaring star into the color range 
we use. For the variation in $M_{r,r-z}$ with magnectic activity and metallicity see, however, the
discussion by \citet{jk21}.  
A more serious contamination can be caused by unresolved binarity and depends on the 
components mass ratio. With a mass ratio of one the colors do not change but the observed $r$ magnitude
is decreased by 0.75 mag. The estimated $M_r$ is not influenced since the $(r-i)$ and $(g-r)$ colors
are left unchanged. For an individual target, binarity introduces a distance uncertainty $\approx$35$\%$
with a unit mass ratio. The estimated extinction is accordingly not altered but its location is shifted
to a larger distance. A recent paper, \citet{jk12} has estimated the fraction of close, a$<$0.4 AU, M 
dwarf binaries to 3$-$4$\%$. Dwarfs later than M6 has a frequency of 20$\pm$4$\%$, \citet{jk13} and the 
overall binary frequency among M dwarfs is 42$\pm$20$\%$, \citet{jk14}.     

In Fig.~\ref{f2} is shown three aspects of the $(r-i) \ vs. \ (g-r)$ diagram for the MBM 12 region sample. 
The black points are a selection the region stars. The green symbols are M dwarfs in the MBM 12
area. Blue symbols signify M dwarfs in the MBM 16 area.
True color excesses are positive but due to observational errors in $(r-i)$ and $(g-r)$ unreddened
and little reddened stars are sometimes shifted to the blue of the $(r-i)_0 \ vs. \ (g-r)_0$ locus.
In Fig.~\ref{f2} we have indicated the one sigma confinement based on the maximum error 0.060 mag in $g, r, i$.
This confinement is corroborated by the scatter around the standard locus of M dwarfs observed at the
virtually unreddened North Galactic Pole.   

\subsection{Estimate of the Extinction $A_r$}

As the sample of nearby M dwarfs, forming the basis for the standard locus, shows there is a scatter
around the standard locus even for vitually unreddened stars. But collapsing the main sequence to a
sharp relation has often proven useful. In a previous section we have argued that a shift from the
locus may mainly depend on reddening. The intrinsic location of a dwarf is on the locus and is determined
by translating the observed position along a reddening vector. The color shifts $\Delta (g-r)$ and
$\Delta (r-i)$ are accordingly assumed to equal $E_{g-r}$ and $E_{r-i}$ respectively.  

$\frac{E_{r-i}}{E_{g-r}}$=0.694 is adopted from \citet{jk04} calculated with $A_V$ = 0.5 mag, $R$=3.1, $T_{eff}$=3500 K, 
$log(g)$=4.5 and $[M/H]$=0. Not
ideal but rather close to the M range. We have $A_r$=2.875$\times$$E_{g-r}$, also adopted from
Girardi et al. op. cit. For $(r-i)$ the relation is $A_r$=4.142$\times$$E_{r-i}$. The ratio between the two
coefficients is 1.441 so other things being equal $E_{g-r}$ is preferred due to a more favorable error 
progression. 

\subsection{Estimated Distance}

Having estimated $A_r$ and observed $r$ only $M_r$ is missing for the photometric parallax.
We rely on the calibration of $M_r$ in terms of $(r-i)_0$ which is preferred to $(g-r)_0$ 
due to a better error progression. 
The $M_r$ calibration of $(r-i)_0$ for M dwarfs is as mentioned adopted from
\cite{jk09}.

The distance estimate is derived from the usual formula:
\begin{equation}
distance = 10^{0.2*(r-M_r-A_r+5)}
\end{equation}

where all parameters are known together with their uncertainties.
\subsection{Uncertainties on Extinction and Distance}

Each entry in the extraction from SDSS DR8 contains errors on all magnitudes but we have preselected
stars with $\sigma_{gri} <$ 0.060 mag. We may estimate the total uncertainty from the 
combination of observational errors and errors introduced from the calibrations.

If $s=f(x_i)$ represents either the full set of equations used
to estimate $A_r$ or $M_r$ the formal error of $s$ follows from the progression formula:

\begin{equation}
 \sigma_s^2 = \Sigma_i (\frac{\partial s}{\partial_{x_i}} \ \sigma_{x_i})^2 
\end{equation}

For each star we compute $\frac{\partial s}{\partial_{x_i}}$ and  $\sigma_{x_i}$ 
implying that derivatives must be calculated and errors of the independent parameters also must be known.
The errors $\sigma_{(g-r)_0}$ and $\sigma_{(r-i)_0}$ are derived from piecemeal linear approximations to
the intrinsic locus, taking into account errors in slope and intersection, together with the observational
errors in $g, r$ and $i$. $\sigma_{M_r}$ is calculated from the calibration equation $(1)$ considering our
individual, estimated errors $\sigma_{(r-i)_0}$. 

The resulting uncertainties are in the range 20$-$26$\%$ for the distances and for $A_r$ about 2/3 has
$\sigma_{A_r}$ in the range from 0.04 to 0.20 mag and about 1/3 between 0.30 and 0.40. The reason for this
double peaked distribution is the kink in the intrinsic locus noticed in Fig.~\ref{f2}

\begin{figure} 
\includegraphics[width=88mm]{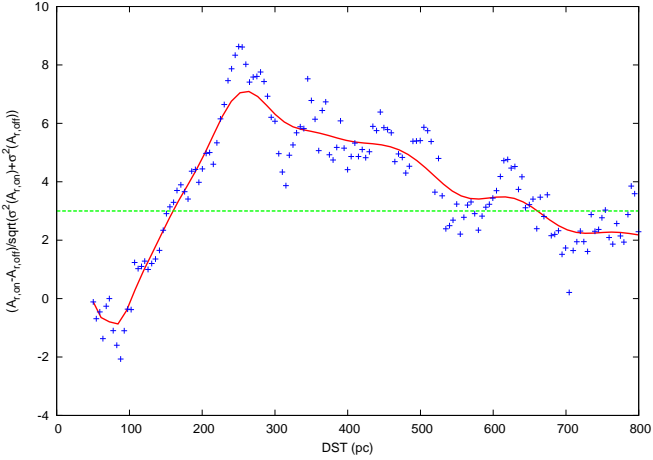}
 \caption{The $\frac{\Delta A_r}{\sigma (\Delta A_r)}$ ratio for the MBM 12 dark cloud.
Points are from 30 pc distance bins, stepsize 10 pc. Neigboring points thus not independent.
The curve is a Bezier smoothing of the data. The ratio equal 3 at 160 pc which is proposed
as the cloud distance} \label{f4}
\end{figure}

\begin{figure} 
\includegraphics[width=88mm]{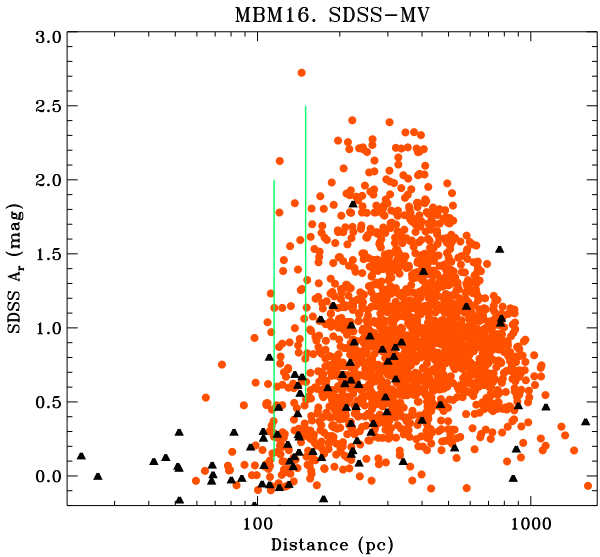}
\caption{Extinctintion $vs.$ distance for the MBM 16 area. Note that beyond $\sim$200 pc 
most stars have $A_r > $0.3 mag. The vertical line indicates that the distance of the first dust is
at 102 pc. The MBM 12 distance at 160 pc is also indicated. Black triangles are Hipparcos data for the
MBM 16 area with $A_r$ extinctions. The star HIP 14997 was left out, it has an 
uncertain $B-V$ and is listed as a variable, \citet{jk20} } \label{f5}
\end{figure}

\begin{figure} 
\includegraphics[width=88mm]{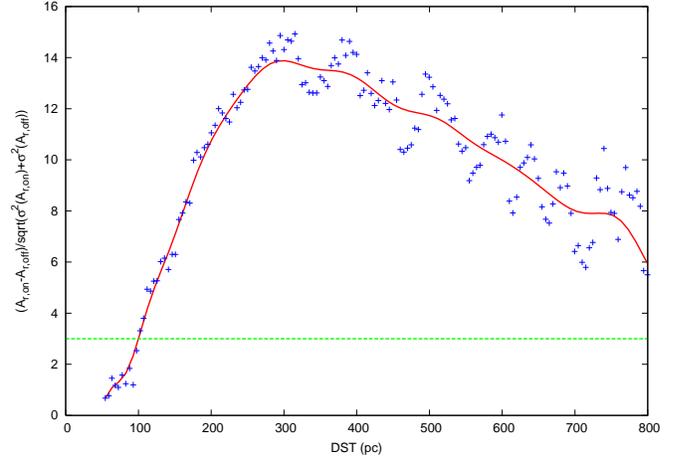}
\caption{The $\frac{\Delta A_r}{\sigma (\Delta A_r)}$ ratio for the MBM 16 translucent 
cloud. The diagram proposes a cloud distance of $\sim$100 pc significantly shorter 
than the $\sim$160 pc suggested for MBM 12} \label{f6}
\end{figure}

\section{Results}

SDSS DR8 provides a substantial amount of M dwarf data for the MBM 12 region and the distance and 
$A_r$ accuracies are adequate to study the distance - extinction variation.

\subsection{MBM 12 Distance Estimate}

The distance to MBM 12 is particularly interesting because the cloud is a specimen of a rare
variety, high latitude, dark cloud ($A_V > 5$ have been measured) that even shows star 
formation activity.

\subsubsection{Distance from appearence of substantial extinction}
By substantial extinction we mean an extinction that appears in a discontinuity and is substantially larger
than the extinction at smaller distances and that several stars do show such an extinction.
 
From the $M_r$ luminosity calibration of $(r-i)_0$ and the color excess from $(g-r)-(g-r)_0$ in Fig.~\ref{f2} we
construct a distance vs. $A_r$ diagram for the MBM 12 area, as shown by the red dots in Fig.~\ref{f3}.
We have used a logarithmic distance scale to emphasize the smaller distances. Apparently there is a rise
in $A_r$ beyond $\sim$0.5 mag around 100 pc. We have shown a vertical line at 125 pc.
There are some MBM 12 stars with extinction $\approx$1 mag at 
$\approx$125 pc but the dominating increase takes place at $\sim$162 pc indicated by a vertical line too.
Since this distance is somewhat smaller than the presently preferred range discussed in the introduction we
have compared to the distance $vs.$ extinction variation derived from the Hipparcos Catalog.
We have extracted Hipparcos stars in the MBM 12 area, same $l$ and $b$ limits as used for the SDSS
extraction. Parallaxes are from the second derivation by \citet{jk15} and extinctions are from $B$,
$V$ photometry and spectral classification, see e.g. \citet{jk16}. The extinction is
averaged in 30 pc intervals and is given as the blue solid curve in Fig.~\ref{f3}. The Hipparcos curve is seen
to follow the upper envelope of the MBM 12 extinction rather well. Since the Hipparcos sample has a 
rather bright limiting magnitude, \citet{jk01}, we can only expect to see smaller extinctions. If we 
accept the distance of the onset of extinctions beyond, say one magnitude, which is $\approx$3$\sigma_{A_r}$above 0, as the cloud distance, MBM 12 is at 160 pc. 

\subsubsection{Alternative distance derivation for MBM 12}

The scans in Fig.~\ref{f1} cover several clouds revealed by their color excesses but also sight lines with less
dust. One may expect that for a small distance bin at a given distance the average extinctions in a cloud
direction and outside the clouds differ. For distances less than the cloud distance, but in the direction
of a cloud, the two averages will be more similar. As Fig.~\ref{f3} shows there is a substantial scatter of 
$A_r$ at almost any distance. A scatter caused by real variation in the presence of dust causing the
extinction and the observational errors, $\sigma_{total}^2=\sigma_{ISM}^2+\sigma_{obs}^2$. If
$\overline{A_{r, on}}$ and $\overline{A_{r, off}}$ designate the average extinctions on and off a cloud
for identical distance bins we propose that $\Delta (A_r)$=$\overline{A_{r, on}}$-$\overline{A_{r, off}}$
will measure the presence of a cloud at a given distance where the difference is sufficiently large. 
We use $\sigma^2 (\Delta(A_r))=\sigma_{total,on}^2+\sigma_{total,off}^2$ as a measure of the significance
of $\Delta (A_r)$ and define the cloud distance as the distance when $\Delta (A_r)$/$\sigma (\Delta(A_r))$
equals three.  

We have done this for the MBM 12 area and compared to the region outside the MBM 12 area as shown in 
Fig.~\ref{f4}. The individual points are not independent: bin size is 30 pc with a steplength of only 10 pc.
The horizontal line is at three and intersects the $\Delta / \sigma$ curve at 160 pc. Which we accept 
as the distance to MBM 12.

\subsection{The distance to MBM 16}

Another cloud in the MBM 12 region is MBM 16 which is of different type than MBM 12. MBM 16 is translucent
whereas MBM 12 is a dark cloud. The two types are distinguished by their optical extinction: a dark
cloud has $A_V > 5$ mag whereas a translucent cloud is less opague with $A_V $ in the range from 1 to 5,
\citet{jk07}. MBM 16 has been studied by \citet{jk17} in order to understand the origin of turbulence and 
the correlations between molecular gas and color excesses.
For their investigation a distance of 100 pc was assumed, identical to the distance to the wall of Local
Bubble as estimated by \citet{jk18}. In their probing of the local low density cavity \citet{jk19} 
suggest the presence of a region with a high density at a distance closer than 100 pc in the direction
of  MBM 16.

In Fig.~\ref{f5} is shown SDSS for what is available inside the area shown in Fig.~\ref{f1}. Hipparcos 2 
results from the same area is overplotted and two vertical lines at 115 and 160 pc respectively. From the
occurence of the first dust, $A_r$ ranging from $\approx$0.5 to $\approx$2 mag, we would suggest $\sim$115 pc.
Smaller than the MBM 12 distance of 160 pc. The $A_r$ data from Hipparcos 2 corroborates this 
to some degree, but only by two stars with $A_r$ in the range from 0.5 to 1 mag.

The $\Delta (A_r)$/$\sigma (\Delta(A_r))$=3 criterion indicates a distance 100 pc for MBM 16 as shown in Fig.~\ref{f6}.

\section{Conclusions}

We have applied two different methods, which perhaps could be termed qualitative and quantitative, respectively,
for estimating the distance to the two high latitude clouds MBM 12 and 16: the distance at which substantial 
extinction is first measured and the distance where the ratio $\Delta (A_r)$/$\sigma (\Delta(A_r))$ equals three.
In the case of MBM 12 and 16 the two methods agree. Our suggested distances are 160 for MBM 12 and 100 pc for 
MBM 16. The former does not agree with the current values from the literature whereas the latter does, almost
exactly. That either method works depend on the nice behaviour of the M2$-$M7 dwarfs in the $(g-r) \ vs. (r-i)$
diagram.   

The difference of the MBM 12 and 16 distances has been narrowed from $\approx$350-100 pc to 160-100 pc. The
possibility that MBM 12 is outside the confinement of the local bubble and MBM 16 is on or inside still exist.
Tempting to suggest that their dark/translucent status is a consequence of their different interstellar environment?

The $\Delta (A_r)$/$\sigma (\Delta(A_r))$ = 3 criterion may possibly be expanded to a generalized method for locating
nearby molecular clouds where $griz$ photometry is available and maps, 2D and 3D, could be produced with a larger
number of stars than used in the spectroscopic study by \citet{b_hli_04}. 

\section*{Acknowledgments}

Our investingation of the Milky Way ISM is financily supported by FNU, grant 09-060601, 
and Fonden af 29. December 1967.

Funding for the Sloan Digital Sky Survey (SDSS) and
SDSS-II has been provided by the Alfred P. Sloan Foundation,
and the Participating Institutions.

\bsp

\label{lastpage}

\end{document}